\documentclass[a4paper,10pt]{article}
\usepackage[utf8]{inputenc}
\usepackage[T1]{fontenc}
\usepackage{microtype}
\usepackage[english]{babel}
\usepackage{amsmath}
\usepackage{amsthm}
\usepackage{algorithmic} 
\usepackage{algorithm}
\usepackage{graphicx}
\usepackage{booktabs}
\usepackage{amsfonts}
\usepackage{wasysym}
\usepackage{textcomp}
\usepackage{tikz}
\usepackage{url}
\usepackage{subfigure}
\usepackage{epstopdf}
\usepackage[toc,page]{appendix}

\makeatletter
\renewcommand*{\@fnsymbol}[1]{\ensuremath{\ifcase#1\or \dagger\or \ddagger\or
   \mathsection\or \mathparagraph\or \|\or **\or \dagger\dagger
   \or \ddagger\ddagger \else\@ctrerr\fi}}
\makeatother

\title{Hydropower optimization: an industrial approach}
\author{Matteo Gardini\thanks{Matteo Gardini is a mathematical engineer graduated at Politecnico of Milan and he works for a2a S.P.A. the second biggest energy company in Italy. matteo.gardini@mail.polimi.it.} , Aurora Manicardi\thanks{Aurora Manicardi is a mathematical engineer graduated at Politecnico of Milan and she works for P\"{o}yry Management Consulting one of the most importat energy strategic consulting companies in Europe.}}
\date{\today}
\begin{document}

\maketitle

\begin{abstract}
Nowadays hydroelectric energy is one of the best energy sources: it is cleaner, safer and more programmable than other
sources. For this reason, its manage could not be done in an approssimative way, but advance mathematical models must be use.
In this article we consider an overview of the problem: we introduce the problem, then we show its simplest 
but quite exaustive mathematical formulation and in the end we produce numerical results under the ipothesis 
that all input are deterministic.
\end{abstract}

\section{Introduction}
During the last fifty years electric energy has become more and more important for the improvement of our society. We use electricity for the majority of our dayly 
activities and a lack of electricity would cause a dive back in the middle age.
An increasing demand of down market energy bought to the introduction of a liberalized market that got of to a keen cimpetition among electic companies. The main goal
is to optimize the electrical energy production, introducing a smart resources management, gas, oil, water, uranium or whatever could be use to generate electicity. An
increasing attention to renewable energies has attached a lot of importance to hydroelectrical energy. This king of energy is substantially clean and perfectly joins the thermoelectrical
programmability t the renewablity giving life to a perfect union that declares it the princess among nowadays energy sources.
\par For this reason such a precious energy can not be menaged in a daredevil way, but it is necessary to use mathematical methods to optimize the use of hydro resource. During the last years several mathematical approach have been developed in order to perform a better hydro sources menagement. These methods involve stochastic and dynamic programming, stochastic and roboust optimization and so on \cite{Korobeinikov2010},\cite{Kristen2009},\cite{Deng2006},\cite{Koopmans1957}. Even if efficient and accurate a lot of methods based on stochastic optimization and programming require an esimation of probability parameters that sometimes could not be so easy to find or maybe obtaining data for estimation may require a lot of money or time. This may happen, for example, if you have to esitmate parameters of flows distribution: you need data from different years, regulary registred in order to find the parameters of time series. And sometimes there is no money, no technology, or, moreoften, no time to do so.\\
For that reason in this article we focus on a simple approach that 
could be use to reach the goal: all you have to know is price and inflows forecast and hydroplant and hydrological basins structural parameters. We derive the corresponding optimization problem, assuming that all variables are deterministic. This problem, althought its semplicity, well explais the principal
variables involved and the most important difficulties that arise in this context. In the first part of the work we introduce the general problem, thn we move to the mathematical formulation and, eventually, we produce a simple application to a real contest, solving the problem numerically.

\section{Problem Overview}
The tipical problem you need to solve when you want to optimise hydroelectrical energy production is, at the fist glance, quite simple. 
Every hydroelectric site is made up of several dreinage basins, such as artificial or natural lakes,  in which flow into hydric counts due to rains, 
tributaries and so on. Downstream of these basins usually a dam is build and then the hydroelectric power plants designeted to electricit production follow. 
These power stations transform water kinetic energy into electicity, using generation groups made up of turbines (such as Kaplan or Francis...) and transformers. 
Every dreinage basin can store a certain amount of water, before it floods, and this water could be use to produce electric energy, whenever economics conditions are favoreble, 
tipically when sell price of energy is higther. Obviously you can't store more water than a baisin could conteins, and you can't produce energy if you haven't got enought water. 
On the other side, every power plant has a minimum amount of energy that have to produce: it could be off, but if it is on he can't produce less then this amount of energy: 
we'll call this quantity of power technical minimum. On the contrary, the maximum power that can be produces it's called maximum handling. The goal of optimization process is to focus 
the energy production those moments in which energy sell prices are higher, satisfying all handling constraints. this problem, 
even if looks simple, is full of pitfalls. To make it harder to solve, could be the complexity of the hydrological site itself, 
that could be enormous and subject to the most different types of constrains. It could happen, for example, that an hydroelectric power plant gets out of order and so could happen that 
would be impossible to produce for a certain laps of time. Moreover a power limitation can occur or maybe a limitation on the volume of the hydrological basin could be imposed. A priori, every little single constrain is able to modify the whole production site management and therefore the result the optimization process. Again, energy power station ability producing could rely on the height of the basin: generally, higher is the level of the lake, higher is the distance of the water from the turbine axle and so greater is the amount of energy produced, on equal water quantity used. And we could carry on with a lot of other examples. All these constrains increase the difficulty of the problem. Moreover, analysis periods are usually long: weeks, months or years and so the number of decisional variables considerably increases. An harder problem raises when constrains or the objective function become non-linear. Even if a lot of smart trick to transform non-linear constrains into linear constrains exist \cite{AIMMS2015}, it is not always possible rewrite a non-linear problem into a linear one. As is known, non-linear problems give life mainly to two king of difficulties:
First, the resolution of a non-linear problem could be very hard and second a global optimum is not guaranteed. 
So, an apparently easy problem has actually a lot of hidden dangers. In this article we mainly focus on linear 
problems, resting sometimes to highlight some situations that could let rise no-linear effects, giving, where is possible, useful linearization advice. 
In the next chapter we show the most general mathematical formulation of the problem and then we move to a linearized form of it.  

\section{Mathematical Formulation}
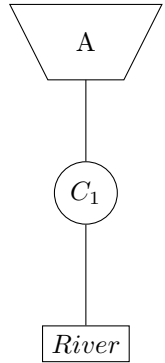
\begin{figure}
\centering
\begin{tikzpicture}

\node at (2.5,4.5) {A};
\draw (2,4) -- (1.5,5) -- (3.5,5) --
	(3,4) -- cycle;

\node [draw,circle] (C1) at (2.5,2.5){$C_1$};
\node [draw] (R) at (2.5,0.5){$River$};
\draw (2.5,4) to (C1) to (R);

\end{tikzpicture}
\caption{Simple hydropower plant.}
\label{Figura_Asta_Semplice_001}
\end{figure}

To formulate the optimization problem in a correct way, it is necessary to understand how a hydroelectric power plant works. Let's consider the simplest case as shown in Figure $\eqref{Figura_Asta_Semplice_001}$.

Generally, upstream there is a drainage basin, $A$, whereas, downstream, the hydroelectric power plant $C_1$ trasforms water cinetic energy into electric energy and discharges water into the river. Let's analise, at the beginning, drainage basin characteristic parameters. 

\subsection{Drainage basin modeling}
A watershed it's characterized by the volume of water that could be stored inside of it. It seems clear that catchement basin main parameters are the maximum and the minimum valume. It may happen that
the maximum volume and the minimum volume change over time: for example, if a certain river drags some bed loads into the lake, the vomum of water will be reduce as the time goes on. Moreover, it could happen that is some periods of the year some landscaping constraints (such as a fixed level of the lake) must hold. Let's define:

\begin{equation}
V_{min}\left(t\right) \quad V_{min}: \mathbb{R} \to \mathbb{R},\quad \quad 
V_{max}\left(t\right) \quad V_{max}: \mathbb{R} \to \mathbb{R},
\nonumber
\end{equation}
maximum and minimum volume functions and
\begin{equation}
V\left(t\right) \quad V: \mathbb{R} \to \mathbb{R},
\nonumber
\end{equation}
that is the volume function, describing the basin volume over time. Let's suppose that $V\left(t\right)$, $V_{min}\left(t\right)$ and $V_{max}\left(t\right)$ are continuos functions, i.e. $V\left(t\right)
\in C\left(\mathbb{R}\right)$.\\ 
Fixing the laps of time $\left[0,T\right]$ , we require that the basin volume is included between the two extreme volumes, i.e.
\begin{equation}
 V_{min}\left(t\right) \le V\left(t\right) \le V_{min}\left(t\right), \quad \forall t \in \left[0,T\right].
 \nonumber
\end{equation}

\par Obviously, the watershed level could be expressed in volumes, measured in $m^3$, or meters above the sea level. For this reason, we could define a particular tipe of curve, called drainage basin curve, which 
curve links volumes and heights. Let's define
\begin{equation}
 \mathfrak{H}\left(h \right) \quad \mathfrak{H}: \mathbb{R} \to \mathbb{R}
 \nonumber
\end{equation}
that connect a given basin height $h$ to the corresponding volume. It is obvious that, for a physical law, this function is at least $C^0$ and, morevore, it is a monotone increasing function. Accordingly, 
$\mathfrak{H}$ has a unique inverse function $\mathfrak{H}^{-1}$ that, given a certain volume $V$, returns the matching height $h$. 
\par Every bowl is feed by the water, which usually flows down from the mountains and flows into the catchement basin, where it can be 
stored using a dam. These incoming discarges must be modeled and we could do it defining another function
that rappresent the amount of water that gets into the watershed over time. Let's state

\begin{equation}
 a\left(t\right) \quad a: \mathbb{R} \to \mathbb{R},
 \nonumber
\end{equation}
which could be a discontinuos function. Clearly, this function can also rappresent rains, snowfalls and all these kind 
of events that could contribute to increase or reduce the water stocked. \\ One of the main arising problems is forecasting these inflows that are uncertain and depending on a lot of exogenous factors \cite{Koskela2009},\cite{Weng2006},\cite{Block2010},\cite{Medeiros2011}.

\subsection{Hydroelectirc power station modeling}
\label{Power_Station}
Let's show now haw can be modeled a hydroelectric power station. A power station is a very complex system
and several mathematical studies has been presented over the years. Anyway, since the objective of
this paper is not to show how model a hydroelectric plant, we can consider it as a black box where water get
in and energy is produced. We can immagine that we have an energy function, that we state with $\mathcal{E}$
\begin{equation}
E=\mathcal{E} \left(\boldsymbol{\theta} \right), \quad \mathcal{E}: \mathbb{R}^n \to \mathbb{R},
\nonumber
\end{equation}
where $E$ is the amount of energy produced in $MWh$,that for a give vector of parameters $\theta$ returns us the amount of energy produced by the plant. Generally speaking,
the amount of energy produced could depend from several factors such that the flow into the turbine $x$, the head $h$ (i.e.
difference in height between the inlet and outlet surfaces),the pressure $p$, the temperature $T$, the efficiency of the turbine
$\eta$ and so on. So we have that $\theta=\left(x,h,p,T,\eta,... \right)$. Modeling of $\theta$ depends on what type of turbine
we consider in our plant. For example if a Pelton turbine is istalled, we must consider that the amount of produced energy
could depends on the head \cite{Catalao2008}, whereas if we have a Francis turbine considering the pressure may be necessary. Anyway, the three
most important variable to consider are, usually, the head $h$, the efficiency $\eta$ and the flow $x$.\\
There are many ways to model the function $E$ but, in practice we can assume that it has the following form:
\begin{equation}
\mathcal{E}=K\left(h,\eta\right)x\left(t\right).
\nonumber
\end{equation}
where $K$ is a coefficient which multiplied by the flow in turbines $x$ gives the energy produced. The expression of $K$
depends on the peculiarity of the problem. The coefficient $K$ may have a very complex form but, in practical purpose,
it could be assumed to be a constant or with a polynomial form like:
\begin{equation}
K\left(h,\eta\right)x\left(t\right)=\eta x\left(t\right) \sum_{j=0}^{m} \gamma_j h^j,
\nonumber
\end{equation}
where $x\left(t\right)$ is the flow at time $t$, $h$ is the head and $\eta$ is turbine's efficiency.

\par Other parameters must be considered, modeling a hydroelectric power plant. Every powerhouse has a own particular range of operation: it can supply a certain quantity of power between a maximum value, that we state with $P_{max}\left(t\right)$, and a minimum value $P_{min}\left(t\right)$. Of  course a generation station can be off. The important point to remind is that if a power station works, then it must dispense power $P\left(t\right)$ such that:
\begin{equation}
P_{min}\left(t\right) \le P\left(t\right) \le P_{max}\left(t\right).
\label{MFPowerConstrain}
\end{equation} 
Commensurating to these two extreme values, expressed in power, there are the corresponding extremity expressed in flow rate $\left( T\left(t\right) \right)$. The constrain $\eqref{MFPowerConstrain}$ is replaced with the following one:
\begin{equation}
T_{min}\left(t\right) \le x\left(t\right) \le T_{max}\left(t\right).
\nonumber
\end{equation} 
This tipe of constraints are called semicontinuos-constraints and could lead to non linear problems. It can be shown that this constraints can be reduced to linear one.\\
Consider a variable $x$ such that $x=0$ or $l \le x \le u$. Choosing a binary variable $y$ we have that the previous one is equivalent to the following ones:
\begin{align}
x \le uy 
\nonumber\\
x\ge ly
\nonumber\\
y \; binary. 
\end{align}
It is clear that $y=0$ implies $x=0$ whereas $y=1$ leads to $l \le x \le u$.

\subsection{Time of concentration}
In the most real cases, the water uses time to flow from a basin to another. For this reason, it is important to model this physical fact. The laps of time that water needs to flow from a place to another is called time of concentration ad we'll state it with $t_{c}$.\\ 
Suppose that water uses a time $t_{c}=1,8$ hours to  move from watershed $A$ to watershed $B$ and consider a time horizon $T$.
Then the volume $V\left(t\right)^B$ is equal to
\begin{equation}
V\left(t\right)^B = V\left(0\right)^B + \int_{0}^{t} a\left(s\right) ds - \int_{0}^{t} x\left(s\right)^B ds + \int_{0}^{t_c} x\left(s\right)^A ds.
\nonumber
\end{equation}
In other words the volume of the drainage basin at time $t$ is given by the starting volume $V\left(0\right)$ plus the inflow less the amount of water released from the basin.

If we moved to discrete time we have to define a discrete grid of time such as
\begin{equation}
0=t_{0} \le t_{1} \le \dots \le T_{k} \le \dots T_{N}=T
\nonumber
\end{equation}
If a certain amount of water $w$ leave the watershed $A$ at time $t_i$, it is easy to that the quantity of water that reaches basin $B$ at time $t_{k+t_c}$ is
\begin{equation}
w \left( 1- \left( t_c - \left\lfloor t_c \right\rfloor \right) \right),
\nonumber
\end{equation}
 while the amount of water at time $t_{k+1+t_c}$ is
\begin{equation}
w \left( t_c - \left\lfloor t_c \right\rfloor \right),
\nonumber
\end{equation}

So, a volume constraint for basin $B$ at time $k$, called $V_{B}^i$, considering time of concentration as above, becomes:
\begin{equation}
V^{B}_{i} = V^{B}_0 + \sum_{i=1}^{k-\left\lfloor t_c\right\rfloor} x_{i}^{A} + \left(1-a\right) x_{\left\lceil t_c\right\rceil} ^{A}\le V^{B}_{max,k}
\nonumber
\end{equation}
where $a=w \left( 1- \left( t_c - \left\lfloor t_c \right\rfloor \right) \right)$ and $x_i$ is the amount of water released from reservoir $A$ at time $i$. 

\subsection{Spillway}
A spillway is a structure used to provide the controlled release of flows from a dam or levee into a downstream area, typically being the river that was dammed. It can be use to avoid incorring in some dangerous situation, for example if the level of the lake becomes to hight.
In continuos time, introducing the spillway, we have that the volume of a general watershed has the following expression:
\begin{equation}
V\left(t\right) = V\left(0\right) + \int_{0}^{t} a\left(s\right) ds - \int_{0}^{t} x\left(s\right) ds - \int_{0}^{t} y\left(s\right)^A ds.
\nonumber
\end{equation}
In our optimization problem, you can choose an amount of water $y\left(t\right)$, that you can throw away in order to avoid to reach a dangerous level. Of course you must minimize this quantity because this lost quantity of water can not be recovered, producing an economic damage. We'll se later how to consider a spillway in our mathematical model. 

\subsection{Objective function}
Let's see now what is the objective function for this kind of problem. Given that every problem has a suitable objective function, the have a general form. Consider, firsply, the continue time problem with a fix time horizon $T$.\\
As stated before we remind that our objective is produce energy where the prices are heigher. So we can consider a function $P_{z}\left(t\right)$ i.e.a price forecasting. $P_{z}\left(t\right)$ could have a very strange form: it could be continuos, discontinuos and there is a lot of letterature about this theme. In our purpose, due to the fact that in the majority of energy market the price of energy in defined hour by hour, we can assume that the function $P_z\left(t\right)$ has the following form:
\begin{equation}
P_{z}\left(t\right)=\sum_{i=0}^{T} p_{i} \mathbb{I}_{\left\{ t \in \left[T_{i-1}, T_{i} \right] \right\}},
\label{Price_Modeling}
\end{equation}
where $\mathbb{I}$ is the indicator function whereas $p_{i}$ are real numbers. So, we can say that, in our purpose, function $P_{z}\left(t\right)$ is a piece-wise constant function.\\ Prices energy forecasting is a crucial point in hydro optimization: a lot of methods have been proposed in order to do so including statistical approach \cite{Khmaladze2007}, econometrics models \cite{Fezzi2007}, arma models \cite{Serati2007} and so on \cite{Gianfreda2011},\cite{Weron2006} but price forecasting is not the aim of this paper.\\
Let'snow  consider the main goal of the article: given an energy price forecasting we want to use the avaiable water in order to maximize the profit that we state with $F\left(\mathcal{E}\left(\boldsymbol{\theta} \right),P_{z}\right)$.
A general expression for $F$ is the following one:
\begin{equation}
F\left( \mathcal{E} \left( \boldsymbol{\theta} \right),P_{z}\right)=\int_{0}^{T} f\left( \mathcal{E} \left( \boldsymbol{\theta} \right),P_{z}\left(s\right)\right) ds
\nonumber
\end{equation}
Suppose that $\theta$ could be split in two sub-vectors suche that
\begin{equation}
\theta=\left[\beta, \alpha \right]
\nonumber
\end{equation}
where $\beta$ rappresents all those parameters that could be choosen (i.e. they are the decison variables), whereas $\alpha$ rappresents the fixed values. The goal is reached choosing those parameters $\beta$ in order to maximize $F$. 
\begin{equation}
\max_{\beta} F\left(\alpha,\beta,P_{z}\right).
\nonumber
\end{equation}
Until now, such a function is very general. Assuming that $\mathcal{E}$ has the simple form presented in
Paragraph $\ref{Power_Station}$ and that $P_{z}$ has the form in Equation $\eqref{Price_Modeling}$ function $F$ assumes the following form:
\begin{equation}
f\left(x\left(t\right),y\left(t\right),\eta,\gamma,h,P_{z}\right)= P_{z}\left(t\right)  \eta \left(x\left(t\right)-y\left(t\right) \right) \sum_{j=0}^{m} \gamma_j h^j.
\nonumber
\end{equation}
In this formulation it's interesting observing how the spillway could be modeled. From an economical point of view, all the water that is splilled is lost and can cause an economical loss. For this reason the term $y\left(t\right) \sum_{j=0}^{m} \gamma_j h^j$ rappresent an economic demage: obviously we have to minimize this term. The sign minus before it forces, during problem resolution, its minimization because in order to maximize the function $F$ we must avoid using $y\left(t\right)$. Actually, $y\left(t\right)$ will be non zero only if a maximum volume constraints would be violated without a zero value for $y\left(t\right)$. From a intuitive point of view it must be clear: I will throw water away if and only if I am not able to store it inside a watershed. It is not necessary that water discharged using the spillway is lost for ever: a hydroplant may have a pumping system which is able to pump water downriver to upriver. As such operation requires electric energy the idea is to pump water when prices are lower and then produce energy when prices are heigher. Such a optimization problem is easy to obtain from the previous one and does not add interesting consideration to our article so we omit it but it can be found in \cite{Korobeinikov2010}. \\


\subsection{The general optimization problem}
Now that we have all the key elements we can show the complete optimization problem.\\
Let's considere a fixed time horizon $T$. Let's consider $N$ watershed and $M$ power plants. According to the notation in previous paragraphes we have that an optimal production strategy can be found solving the following optimization problem:

\begin{equation}
\max_{\boldsymbol{\beta}} F\left( \mathcal{E} \left( \boldsymbol{\beta},\boldsymbol{\alpha} \right),P_{z}\right) =\int_{0}^{T} f\left( \mathcal{E} \left( \boldsymbol{\theta} \right),P_{z}\left(s\right)\right) ds
\nonumber\\
\end{equation}
subject to

\begin{align}
V_{min}^{i}\left(t\right) & \le V^{i}\left(t\right) \le  V_{max}^{i}\left(t\right) \quad & i=1,\dots,N.
\nonumber\\
\max_{j \in down_i} T_{min}^j\left(t\right) & \le  x^i \left(t\right) \le \min_{j \in down_i} T_{max}^j \left(t\right) \quad & i =1,\dots,N \; j=1,\dots,M,
\nonumber
\end{align}
where $down_i$ is the set of power plants which are aligned under the same basin $i$. This constraints may seem obscure but is extreamly simple to understand: if we have two hydropower plants $A$ on the top and then, immediatly downriver, plant $B$ and we have $T_{max}^{A}=\delta$ and $T_{max}^{B}=\zeta$ with $\delta > \zeta$ then it's obvious that the maximum amount of water that can be released from the basin on the peack is $T_{max}^{B} = \max_{j \in \left(A,B\right)} T_{max}^{j}\left(t\right)$. An analogous reasoning holds for $T_{min} \left(t\right)$.

\par In order to solve the problem numerically, it's suitable to move to discrete space so that the problem can be formulated as a typical operational research problem.\\
Let's firstly consider a grid partition of the time period $\left[0,T\right]$: this grid is equispaced grid with $N_h$ points, where each point rappresent an hour. This assumption is reasonable because, for example, in day haed marked we have hourly prices and so it's common use deciding hour by hour how much energy producing. Anyway it is possible to fix an an arbitrarily fine grid. So, our grid is of the following form:
\begin{equation}
0=t_0 \le \dots \le t_k \le \dots \le t_{N_h}=T.
\nonumber
\end{equation}
As observed above, every bowl in every discrete instant is feed by some tributaries, that bring a certain amount of water, that we state with $a_{k}^i$ where $i=1,\dots,N$ and $k=1,\dots,N_h$.
Let's state with $V_{k}^i$ the volume of the lake $i$ at time $k$, with $T_{max,k}^j$ the maximum amount of water that can pass through the power station $j$ at time $k$ and with $T_{min,k}^j$ the
corresponding minimum. \\
The next step is to discretize the objective function: appling a reasoning like above, we have the following form:

\begin{equation}
\max_{\boldsymbol{\beta}} F\left( \mathcal{E} \left( \boldsymbol{\beta},\boldsymbol{\alpha} \right),P_{z}\right)= \max_{\beta} \sum_{k=1}^{N_h} \sum_{i=1}^{N} \sum_{j \in M \cap  down_i} f_{j}\left( \mathcal{E}_{j} \left( \boldsymbol{\beta}_{i},\boldsymbol{\alpha} \right),P_{z_{i}}\right) 
\nonumber
\end{equation}


We are ready now to formulate the complete optimization problem.
Under the assumptions above a general problem can be formulate in the following way. 

\begin{equation}
\max_{\boldsymbol{\beta}} F\left( \mathcal{E} \left( \boldsymbol{\beta},\boldsymbol{\alpha} \right),P_{z}\right)=\max_{\beta} \sum_{k=1}^{N_h} \sum_{i=1}^{N} \sum_{j \in M \cap  down_i}  f_{j}\left( \mathcal{E}_{j} \left( \boldsymbol{\beta}_{i},\boldsymbol{\alpha} \right),P_{z_{k}}\right)\nonumber\\
\end{equation}
subject to
\begin{align}
V_{min,k}^i \le  V_{k}^i & \le V_{max,k}^i & i=1,\dots,N \; k=1,\dots, N_h 
 \nonumber\\
 \max_{j \in down_i} T_{min,k}^j  & \le  x_{k}^i \le \min_{j \in down_i} T_{max,k}^j & i =1,\dots,N \; k=1,\dots, N_h,
 \nonumber
\end{align}
where $down_i$ is the set of power plants depending by the water released by the basin $i$ and where $P_{z_k}$ is the sell price of
energy at time $k$.\\
This kind of problem could be linear or not, it depends on the form of objective function $f$ and on the nature of constraints.
Generally, volumes constraints are linear while constraints about maximum and minimum water could be harder to write and leads to
a interger optimization problem. Anyway, in the simplest case such a problem could be reduce to a linear optimization problem,
assuming a constant $ f_{j}\left( \mathcal{E}_{j} \left( \boldsymbol{\beta}_{i},\boldsymbol{\alpha} \right),P_{z_{k}}\right)\nonumber$ for all $j=1,\dots,M$, for all $i=1,\dots,N$ and for all $k=1,\dots,N_h$. Moreover we assume that $T_{min,k}^j=0$ for all $i=1,\dots,N$ and $k=1,\dots,N_k$. Let's observe that this
assumption seems to be very strong but, in practice, leads to good solution. The reason of this fact is that, generally, hydroelectic
power plants has a $T_{min}$ that is very low (sometime equal to $2$ or $3\frac{m^3}{s}$ ore even less). Moreover, the particular shape of prices outline, in these
kind of problem, with only rare exceptions, bring power plants to work at the maximum power or to be off. Rarely happens that a power station works at a $x_{k}^i$ that is near zero. Anyway, this must not be considered a rigorous assumption, shall be non optimality of solution. If you want to be more accurate you have to introduce semicontinuos variable as we shown above at the end of Paragraph $\ref{Power_Station}$. 
\par In the next chapter, we consider a real case and we write all constraints explicitly in order to make the dissertation clearer.

\section{Numerical Results}
\begin{figure}
\begin{tikzpicture}

\node at (2.5,13.5) {A};
\draw (2,13) -- (1.5,14) -- (3.5,14) --
       (3,13) -- cycle;

\node at (8.5,13.5){B};
\draw (8,13) -- (7.5,14) -- (9.5,14) --
       (9,13) -- cycle;

\node at (5.5,7.5){C};
\draw (5,7) -- (4.5,8) -- (6.5,8) --
       (6,7) -- cycle;

\node at (11.5,10.5){D};
\draw (11,10) -- (10.5,11) -- (12.5,11) --
       (12,10) -- cycle;

\node at (8.5,3.5){E};
\draw (8,3) -- (7.5,4) -- (9.5,4) --
       (9,3) -- cycle;

\node [draw,circle] (C1) at (2.5,11.5){$C_1$};
\node [draw,circle] (C2) at (2.5,9.5){$C_2$};
\node [draw,circle] (C7) at (8,12){$C_7$};
\node [draw,circle] (C3) at (9,12){$C_3$};
\node [draw,circle] (C4) at (6.5,5.5){$C_4$};
\node [draw,circle] (C5) at (11.5,5.5){$C_5$};
\node [draw,circle] (C6) at (8.5,1.5){$C_6$};
\node [draw] (River) at (4,1.5){River};

\draw (2.5,13) to  (C1);
\draw (C1) to  (C2);
\draw (C2) to (5.5,9.5) to (5.5,8) ;
\draw (9,13) to (C3) to (11.5,12) to (11.5,11);
\draw (8,13) to (C7) to (5.5,12) to (5.5, 8);
\draw (11.5,10) to (C5) to (8.5,5.5) to (8.5,4);
\draw (5.5,7) to (5.5,5.5)  to (C4) to (8.5,5.5) to (8.5,4);
\draw (8.5,3) to (C6) to (River);
\draw [dashed] (9.5,14) to (14,14) to (14,4) to (9.5,4); 
\draw [dashed] (4.5,8) to (4,8) to (River);

\end{tikzpicture}
\caption{A complex hydropower plant}
\label{asta_idroelettrica}
\end{figure}
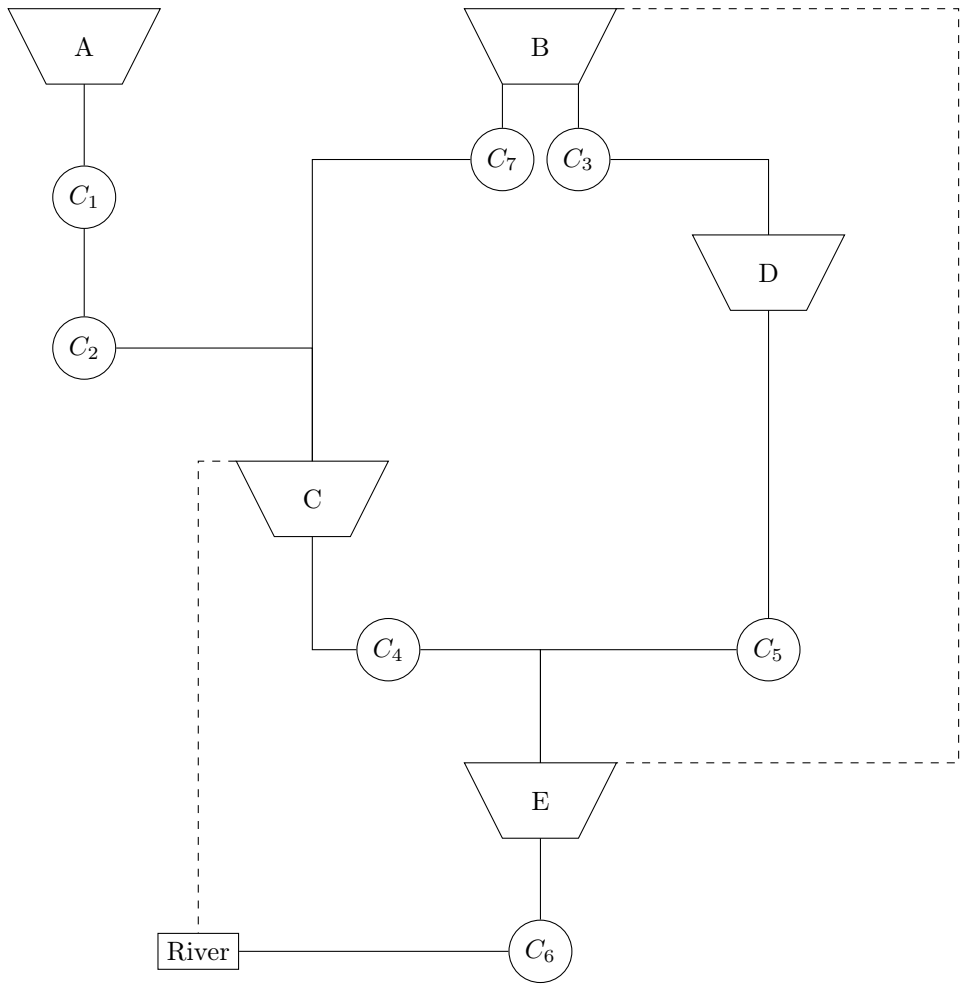

This chapter provides numerical solution of the problem descrived before. $MATLAB \textregistered$ is used and all simulation have been performed on a Desktop PC running Ubuntu 12.04 
with a processor AMD Athlon(tm2)X4, CPU 2.60Ghz and whit a RAM of 3.25 GB.\\ 
Let's consider the hydro-power site shown in Figure $\eqref{asta_idroelettrica}$.

In this picture there are $N=5$ watershed, named with letters from $A$ to $E$ and $M=7$ power plants named with numbers from $1$ to $7$. 
Continuos lines rappresent real pipes where water is transported, dotted lines shall be considered as possible spills. A certain amount of water, stated with $a^i$, 
flows into each basin. All power plants has a minimum amount of water that they can process $T_{min}^j$  and a maximum one $T_{max}^j$. 
Moreover, any hydropower plants has a specific function $K_j\left(h_i,\eta\right)$ which rapresents energy produced from a certain amount of water. 
Let's consider a time horizon of one day and a hourly time stem. $\pi_{k}$ will rappresent the value of the generic quantty $\pi$ at discrete time $k$.\\
Let's assume some time of concentration, expressed in hours, as follow: $t_{CE}=1$, $t_{BD}=1$, $t_{DE}=1$ $t_{BE}=2$ and $t_{CE}=1$: the otherones will be setted to zero.
\par As previously explained we assume $T_{min}^{j}=0 \; \forall j=1,\dots,M$.\\
Let's now define some parameters of the model: starting from the basins, they are characterized by volumes shown in Table $\eqref{Tabella_Volumi_1}$ (assumed constant
over the time). 

\begin{table}[!h]
\centering
\begin{tabular}{cccc}
\toprule
$V_{min}$ & $V_{max}$  & $V_{0}$ & $V_{T}$\\
$m^3$     & $m^3$   & $m^3$ & $m^3$\\
\midrule
1.2 & 60 & 5 & 5 \\ 
2.5 & 40 & 5 & 5 \\ 
3.2 & 55 & 7 & 7 \\ 
3.1 & 70 & 5 & 5 \\ 
4.2 & 90 & 20 & 20 \\ 
\bottomrule
\end{tabular}
\caption{Watershed volumes.}
\label{Tabella_Volumi_1}
\end{table}

Moreover they have hourly inflows ($a_{k}^i \; i=1,\dots,N \; k=1,\dots,M$) showed in Table $\eqref{Tabella_Apporti_1}$.

\begin{table}[!ht]
\centering
\begin{tabular}{ccccc}
\toprule
$a^{A}$ 	  & $a^{B}$    & $a^{C}$  & $a^{D}$  & $a^{E}$\\
$m^3/h$      & $m^3/h$   & $m^3/$ & $m^3/h$  & $m^3$/h\\
\midrule
 2.3337 & 3.0221 & 2.8945 & 3.9531 & 2.6138 \\ 
 2.2028 & 3.5317 & 2.3827 & 4.5236 & 4.2454 \\ 
 4.2818 & 4.6967 & 4.2817 & 3.4773 & 3.0284 \\ 
 2.9336 & 2.0453 & 2.5520 & 4.7758 & 2.5519 \\ 
 4.4519 & 2.6908 & 2.9745 & 2.8622 & 3.6579 \\ 
 2.4645 & 4.0023 & 3.7028 & 3.4258 & 4.8324 \\ 
 3.2934 & 4.2211 & 3.3939 & 4.2016 & 3.4684 \\ 
 2.7164 & 4.4360 & 2.6784 & 2.1096 & 2.7181 \\ 
 4.3255 & 3.4349 & 2.4897 & 3.3208 & 4.2712 \\ 
 2.0225 & 4.2963 & 2.2958 & 2.2968 & 2.7570 \\ 
 3.6610 & 3.3393 & 3.4493 & 4.1444 & 3.2979 \\ 
 2.7418 & 4.2315 & 3.2741 & 3.2971 & 3.3728 \\ 
 3.2016 & 2.9761 & 3.7503 & 4.2663 & 2.0503 \\ 
 3.0557 & 4.8328 & 4.8110 & 4.4548 & 4.6104 \\ 
 2.7563 & 2.9621 & 3.3225 & 4.8770 & 4.2411 \\ 
 2.3141 & 4.2988 & 4.1209 & 2.0497 & 3.2352 \\ 
 3.2254 & 3.8098 & 4.7184 & 3.4141 & 3.5946 \\ 
 3.5101 & 2.4082 & 3.4361 & 2.0051 & 4.4080 \\ 
 2.0871 & 4.6195 & 4.9013 & 4.1003 & 2.7787 \\ 
 2.8884 & 3.2551 & 3.2188 & 4.0456 & 3.7858 \\ 
 4.8182 & 3.9877 & 4.5096 & 2.0498 & 4.0051 \\ 
 2.1249 & 4.0510 & 4.4390 & 2.4314 & 4.0160 \\ 
 2.8717 & 2.0767 & 4.3196 & 3.1859 & 4.4756 \\ 
 3.1643 & 3.8879 & 3.5245 & 4.3220 & 3.4821 \\ 
\bottomrule
\end{tabular}
\caption{Inflows.}
\label{Tabella_Apporti_1}
\end{table}

Each power plant has technical parameters shown in Table $\eqref{Tabella_Centrali_1}$.

\begin{table}[!ht]
\centering
\begin{tabular}{ccc}
\toprule
$j$ & $K_{j}$ & $T_{max}^j$ \\	
$-$ & $kWh/m^3$ & $m^3/h$ \\
\midrule
1 & 0.8 & 15 \\
2 & 0.9 & 25 \\
7 & 1.1 & 15 \\
3 & 1.1 & 25 \\
4 & 1.2 & 15 \\
5 & 3.4 & 85 \\
6 & 1.2 & 35 \\
\bottomrule
\end{tabular}
\caption{Power plants parameters.}
\label{Tabella_Centrali_1}
\end{table}

Last but not least we need a price forecast: an example is shown in Table $\eqref{Tabella_Prezzi_1}$.

\begin{table}[!ht]
\centering
\begin{tabular}{cccc}
\toprule
$Hour$ & $Prices$ & $Hour$ & $Prices$\\
$-$ & $EUR/MWh$ & $-$& $EUR/MWh$\\
\midrule
1 & 10.11 & 13  & 11.13\\
2 & 12.45 & 14 &  11.13\\
3 & 12.45 & 15 & 11.13\\
4 &14.11 & 16 & 11.15\\
5 & 20.34 & 17 & 11.15\\
6 & 25.78 & 18 & 21.34\\
7 & 40.44 & 19 & 34.45\\
8 & 55.45 & 20 & 45.55\\
9 & 41.76 & 21 & 88.90\\
10 & 20.01 & 22 & 88.90\\
11 & 10.04 & 23 & 32.33\\
12 & 12.45 & 24 & 10.11\\
\bottomrule
\end{tabular}
\caption{Inflows.}
\label{Tabella_Prezzi_1}
\end{table}

\noindent Once we wrote all constraints and the objective function to maximize we are ready to solve numerically the problem.\\
Using $MATLAB \textregistered$, in particular the $linprog$ solver, we have water releases from the drainage basins, shown in Table $\eqref{Tabella_Rilasci_Linear_1}$.

\begin{figure}[!ht]
\centering
\includegraphics[scale=0.7]{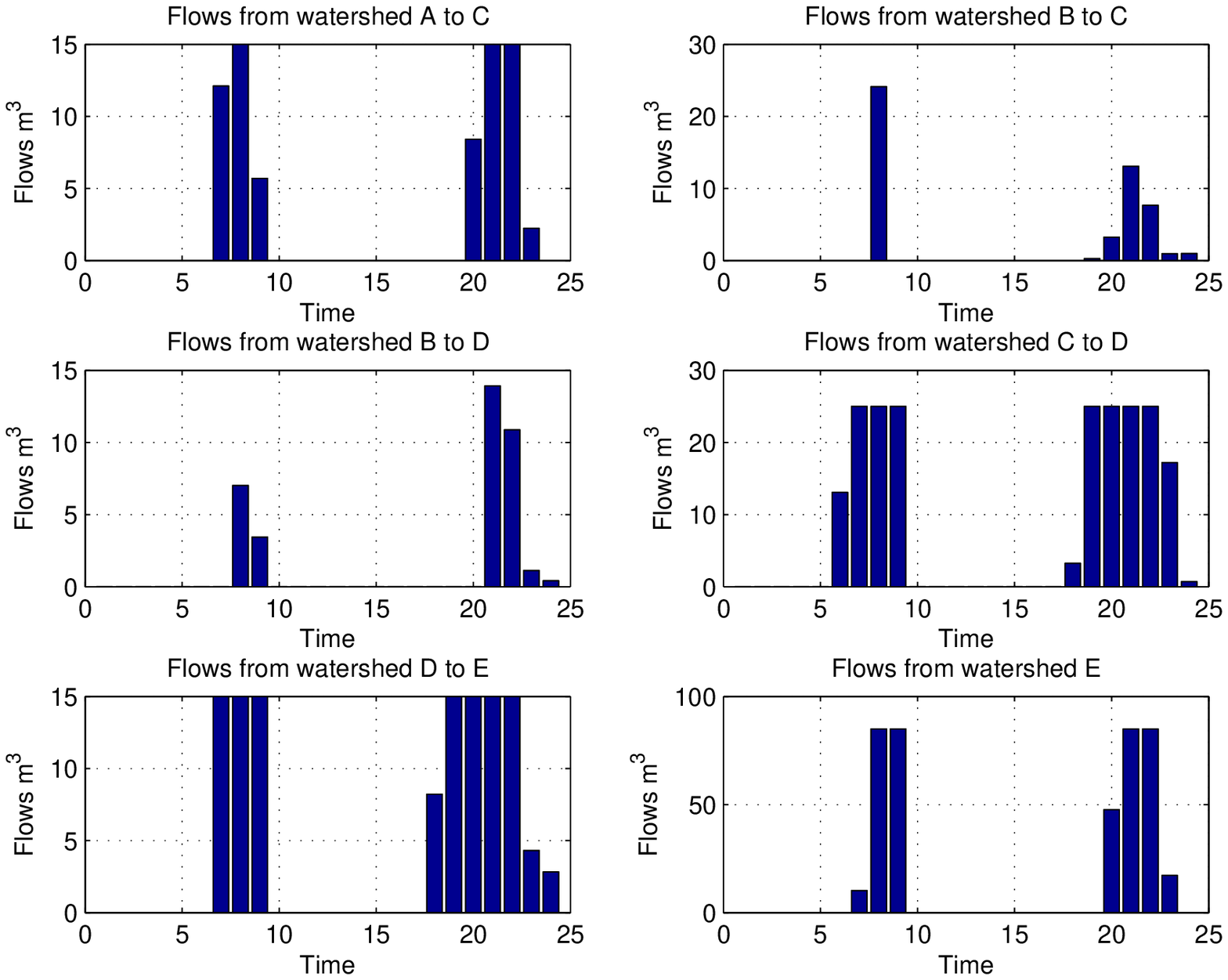}
 \label{Turb_Lin_1}
 \caption{Productions in linear case.}
\end{figure}

\begin{table}[!ht]
\centering
\begin{tabular}{cccccccc}
\toprule
$Rel^{AC}$ 	  & $Rel^{BC}$    & $Rel^{BD}$  & $Rel^{CE}$  & $Rel^{DE}$ & $Rel^{ER}$ & $Spill^{CR}$ & $Spill^{BE}$\\
$m^3/h$      & $m^3/h$   & $m^3/h$ & $m^3/h$  & $m^3/h$ & $m^3/h$  & $m^3/h$ & $m^3/h$\\
\midrule
 0.0000 & 0.0000 & 0.0000 & 0.0000 & 0.0000 & 0.0000 & 0.0000 & 0.0000 \\ 
 0.0000 & 0.0000 & 0.0000 & 0.0000 & 0.0000 & 0.0000 & 0.0000 & 0.0000 \\ 
 0.0000 & 0.0000 & 0.0000 & 0.0000 & 0.0000 & 0.0000 & 0.0000 & 0.0000 \\ 
 0.0000 & 0.0000 & 0.0000 & 0.0000 & 0.0000 & 0.0000 & 0.0000 & 0.0000 \\ 
 0.0000 & 0.0000 & 0.0000 & 0.0000 & 0.0000 & 0.0000 & 0.0000 & 0.0000 \\ 
 0.0000 & 0.0000 & 0.0000 & 13.0862 & 0.0000 & 0.0000 & 0.0000 & 0.0000 \\ 
 12.1037 & 0.0000 & 0.0000 & 25.0000 & 15.0000 & 10.2743 & 0.0000 & 0.0000 \\ 
 15.0000 & 24.1314 & 7.0149 & 25.0000 & 15.0000 & 85.0000 & 0.0000 & 0.0000 \\ 
 5.7003 & 0.0000 & 3.4349 & 25.0000 & 15.0000 & 85.0000 & 0.0000 & 0.0000 \\ 
 0.0000 & 0.0000 & 0.0000 & 0.0000 & 0.0000 & 0.0000 & 0.0000 & 0.0000 \\ 
 0.0000 & 0.0000 & 0.0000 & 0.0000 & 0.0000 & 0.0000 & 0.0000 & 0.0000 \\ 
 0.0000 & 0.0000 & 0.0000 & 0.0000 & 0.0000 & 0.0000 & 0.0000 & 0.0000 \\ 
 0.0000 & 0.0000 & 0.0000 & 0.0000 & 0.0000 & 0.0000 & 0.0000 & 0.0000 \\ 
 0.0000 & 0.0000 & 0.0000 & 0.0000 & 0.0000 & 0.0000 & 0.0000 & 0.0000 \\ 
 0.0000 & 0.0000 & 0.0000 & 0.0000 & 0.0000 & 0.0000 & 0.0000 & 0.0000 \\ 
 0.0000 & 0.0000 & 0.0000 & 0.0000 & 0.0000 & 0.0000 & 0.0000 & 0.0000 \\ 
 0.0000 & 0.0000 & 0.0000 & 0.0000 & 0.0000 & 0.0000 & 0.0000 & 0.0000 \\ 
 0.0000 & 0.0000 & 0.0000 & 3.2368 & 8.2167 & 0.0000 & 0.0000 & 0.0000 \\ 
 0.0000 & 0.2749 & 0.0000 & 25.0000 & 15.0000 & 0.0000 & 0.0000 & 0.0000 \\ 
 8.4078 & 3.2552 & 0.0000 & 25.0000 & 15.0000 & 47.6071 & 0.0000 & 0.0000 \\ 
 15.0000 & 13.0788 & 13.9104 & 25.0000 & 15.0000 & 85.0000 & 0.0000 & 0.0000 \\ 
 15.0000 & 7.6762 & 10.8732 & 25.0000 & 15.0000 & 85.0000 & 0.0000 & 0.0000 \\ 
 2.2361 & 0.9491 & 1.1277 & 17.2085 & 4.3137 & 17.2146 & 0.0000 & 0.0000 \\ 
 0.0000 & 0.9755 & 0.4125 & 0.7000 & 2.8345 & 0.0000 & 0.0000 & 0.0000 \\ 
\bottomrule
\end{tabular}
\caption{Results of linear problem.}
\label{Tabella_Rilasci_Linear_1}
\end{table}
As expected we can observe that the production in higher when prices are higher. Moreover, due to the fact that the water inflows are low, the spills are null. 
There's no reason to throw water away in this case: it is more conveniente turbine water and produce energy. The solution of this extrimely simple lineare problem 
request only $2.5431$ seconds.
Even if this is a quite simple problem, this solution it's better than the one we would have found tring to programming the plant using common sense or experience: that is
due to the fact that mathematical theorems guarantie that the solution is optimal. 
In Figure $\eqref{Turb_Lin_1}$ it clearer that production is focus where energy prices are heigher.\\

Next step is to consider a non linear case where all the parameters remain the previously defined except for the $K_{e}$ function which is now non constant.
Let's suppose that the efficiency of the plants depends on the height of the level of the lake. This assumption is realistic if we consider Pelton turbines which 
are used when the water head is hight and the flow rate is small. Among the various possibilities to model the $K_{e}$ a polinomial function is considered: 
$K_{e}=\gamma_{0} + \gamma_{1} x_{1} +\gamma_{2} x_{2}^2$ with parameters shown in Table $\eqref{Tabella_Centrali_2}$. 

\begin{table}[!ht]
\centering
\begin{tabular}{cccc}
\toprule
$j$ & $\gamma_{0}$ & $\gamma_{1}$ & $\gamma_{2}$ \\	
\midrule
1 & 0.80 &	1.170e-01 &	-4.229e-03 \\
2 & 0.9 & 	2.170e-01 &	-2.229e-03\\
7 & 1.2 & 	3.170e-01 &	-1.229e-03\\
3 & 1.1 & 	2.170e-01 &	-3.229e-03\\
4 & 1.1 &	0.170e-01 &	-7.229e-03\\
5 & 1.2 & 	4.170e-01 &	-3.229e-05\\
6 & 3.4 &	2.170e-01 &      -7.229e-03\\
\bottomrule
\end{tabular}
\caption{Power plants parameters in non-linear case.}
\label{Tabella_Centrali_2}
\end{table}

The amount of energy produced at time $t$ depends on the head of the upstream basin, which depends on the amount of water released at $t-1$.
The polynomial form of the $K_{e}$ leads to a non linear objective function which can sometimes be linearized thought specific linearization techniques.
In this case constraints are still linear but a linearization approach is always suggested in all other situations: linear problems can be solved faster, easier trough 
a lot of numerical methods which lead to a global solution.
Anyway, in order to show how to solve a non linear problem and show a lack of efficiency in finding solution, we do not linearize the objective function: we solve the 
problem using $MATLAB \textregistered$ $fmincon$ solver. Results are shown in Table $\eqref{Tabella_Rilasci_NON_Linear_1}$.

\begin{table}[!ht]
\centering
\begin{tabular}{cccccccc}
\toprule
$Rel^{AC}$ 	  & $Rel^{BC}$    & $Rel^{BD}$  & $Rel^{CE}$  & $Rel^{DE}$ & $Rel^{ER}$ & $Spill^{CR}$ & $Spill^{BE}$\\
$m^3/h$      & $m^3/h$   & $m^3/h$ & $m^3/h$  & $m^3/h$ & $m^3/h$  & $m^3/h$ & $m^3/h$\\
\midrule
 0.0000 & 0.0000 & 0.0000 & 0.0000 & 0.0000 & 0.0000 & 0.0000 & 0.0000 \\ 
 0.0000 & 0.0000 & 0.0000 & 0.0000 & 0.0000 & 0.0000 & 0.0000 & 0.0000 \\ 
 0.0000 & 0.0000 & 0.0000 & 0.0000 & 0.0000 & 0.0000 & 0.0000 & 0.0000 \\ 
 0.0000 & 0.0000 & 0.0000 & 0.0000 & 0.0000 & 0.0000 & 0.0000 & 0.0000 \\ 
 0.0000 & 0.0000 & 0.0000 & 0.0000 & 0.0000 & 0.0000 & 0.0000 & 0.0000 \\ 
 0.0000 & 0.0000 & 0.0000 & 0.0000 & 0.0000 & 0.0000 & 0.0000 & 0.0000 \\ 
 0.0000 & 0.0000 & 10.6798 & 25.0000 & 0.5876 & 0.0000 & 0.0000 & 0.0000 \\ 
 10.7081 & 0.0000 & 15.0000 & 13.0033 & 15.0000 & 85.0000 & 0.0000 & 0.0000 \\ 
 6.1392 & 0.0000 & 5.4900 & 9.9946 & 15.0000 & 32.4525 & 0.0000 & 0.0000 \\ 
 0.0000 & 0.0000 & 0.0000 & 2.2958 & 0.0000 & 0.0000 & 0.0000 & 0.0000 \\ 
 0.0000 & 0.0000 & 0.0000 & 0.0000 & 0.0000 & 0.0000 & 0.0000 & 0.0000 \\ 
 0.0000 & 0.0000 & 0.0000 & 6.7235 & 0.0000 & 0.0000 & 0.0000 & 0.0000 \\ 
 0.0000 & 0.0000 & 0.0000 & 1.4532 & 0.0000 & 0.0000 & 0.0000 & 0.0000 \\ 
 0.0000 & 0.0000 & 0.0000 & 0.0000 & 0.0000 & 0.0000 & 0.0000 & 0.0000 \\ 
 0.0000 & 0.0000 & 0.0000 & 0.0000 & 0.0000 & 0.0000 & 0.0000 & 0.0000 \\ 
 0.0000 & 0.0000 & 0.0000 & 0.0000 & 0.0000 & 0.0000 & 0.0000 & 0.0000 \\ 
 0.0000 & 0.0000 & 0.0000 & 0.0000 & 0.0000 & 0.0000 & 0.0000 & 0.0000 \\ 
 0.0000 & 0.0000 & 0.0000 & 0.0000 & 3.1480 & 0.0000 & 0.0000 & 0.0000 \\ 
 0.0000 & 0.0000 & 3.6863 & 25.0000 & 15.0000 & 85.0000 & 0.0000 & 0.0000 \\ 
 15.0000 & 3.2551 & 0.0000 & 24.0814 & 15.0000 & 0.0000 & 0.0000 & 0.0000 \\ 
 15.0000 & 6.2303 & 15.0000 & 25.0000 & 15.0000 & 85.0000 & 0.0000 & 0.0000 \\ 
 15.0000 & 4.8209 & 12.4844 & 25.0000 & 15.0000 & 50.6493 & 0.0000 & 0.0000 \\ 
 11.6006 & 9.0797 & 0.0000 & 25.0000 & 15.0000 & 45.2761 & 0.0000 & 0.0000 \\ 
 0.0000 & 1.3879 & 0.0000 & 1.1125 & 15.0000 & 9.5217 & 0.0000 & 0.0000 \\ 
\bottomrule
\end{tabular}
\caption{Results of non linear problem.}
\label{Tabella_Rilasci_NON_Linear_1}
\end{table}

\begin{figure}[!h]
\centering
\includegraphics[scale=0.7]{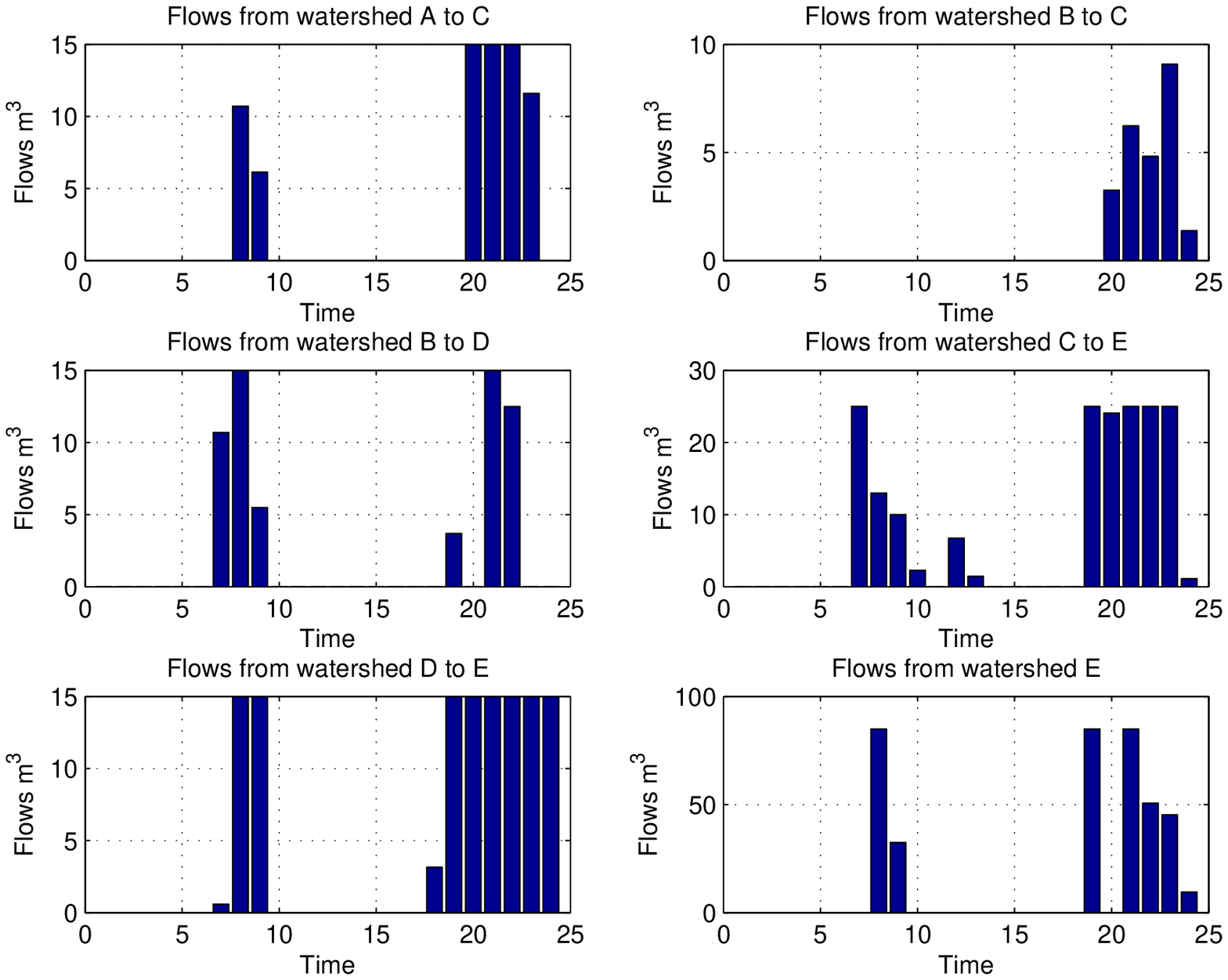}
 \label{Turb_NON_Lin_1}
 \caption{Productions in non linear case.}
\end{figure}

Again, in Figure $\eqref{Turb_NON_Lin_1}$ productions in the non-linear case are shown. We can see that the results are quite similar to the previous experiment. 
The differences beetween linear and non-lineare case depend on the difference of energetic coefficients. The solution of the non-linear problem tooks $11.14$ seconds, more
than the linear case, as one should expect.

\begin{figure}[!h]
\centering
\includegraphics[scale=0.7]{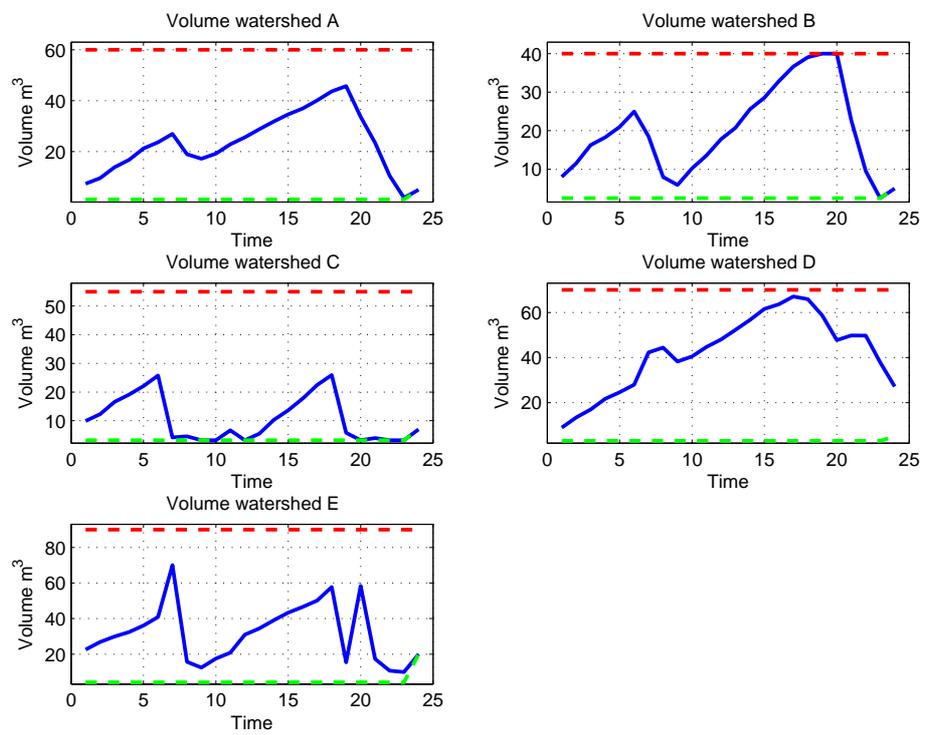}
 \label{Volumes_NON_Linear_1}
 \caption{Volumes. At the end of the period we have that all volumes reach the designed values $V_{T}^i$.}
\end{figure}

Let's have a look to the volumes in Figure $\eqref{Volumes_NON_Linear_1}$. 
As we can see all the volumes constraints are respected: indeed the volumes of each watershed lies between $V_{min}^i$ and $V_{max}^i$. 
Moreover, by the end of period, at time $T$ we have that the volumes of watershed $i$ has reached the value of $V_{T}^i$. 
We have to observe that we have not request that $V_{N}^i=V_{T}^i$ but only that $V_{N}^i \ge V_{T}^i$. 
Looking at problem formulation should be clear that all possible water will be use to produce energy because producing energy maximize the objective function: 
however, if not all water can be processd but could be stored into the basin because the maximum volume has not been reached yet, then we might have 
$V_{N}^i \ge V_{T}^i$. 

As last consideration we can observe that if we have that the efficiency of the turbine depends on the head it may happen that even if the price is 
good it may be more convenient store water in order to increase the head and the efficiency of the plant. Then we will be able to produce the same amount of energy using
less water, taking advantage of the heigher level of the basin.
\clearpage


\section{Conclusions}
In this paper we focused on short-term hydroelectric modeling and optimization. If we want to develop
a long-term optimization model the problem is the same \cite{Zhao2011}. Same constraints, same objective function, same
input and output. One of the main problem that arise wher we develop a long-term model is that the complexity
of the problem quickly increases. In these cases find a optimal solution for the problem might request a
long time. Softwares that are able to solve \textit{large-scale} problems are avaiable and must be used in
to solve this kind of problems. They includes $LINGO$ \cite{LINGO2015}, $AMPL$ \cite{AMPL2015} $CPLEX$ \cite{CPLEX2015}.\\
Moreover we have considered only deterministic input values. Actually, all input variables such as inflows,
price forecast are not deteministic but stochastic. If one is able to stimate the statistical destribution of
this quantity a stochastic optimization approach could be used such as in \cite{Pereira2006},\cite{Kristen2009},\cite{Deng2006}. If this probability distribution could not be infer a robous optimization should be chosen.\\
As observed before, evenif its semplicity, this appoach could be very successfull from an industrial point of view. Generally all input needed are easy to find. Moreover the mathematical approach is very standard and easy to understand whereas the numerical problem can be implemented using some very userfriendly softwares such as $What'sBest!$ \cite{WB2015} which provides an easy $Excel$ interface that is very simple to use for a low-medium user and well documented. Furthermore this approach could be studied and implemented from a single person and this could be very useful for those companies which are small or do not have a specialized research and development department.

\begin{appendices}
	\section{Appendix A}
\begin{figure}[!ht]
	\centering
	\includegraphics[scale=0.5]{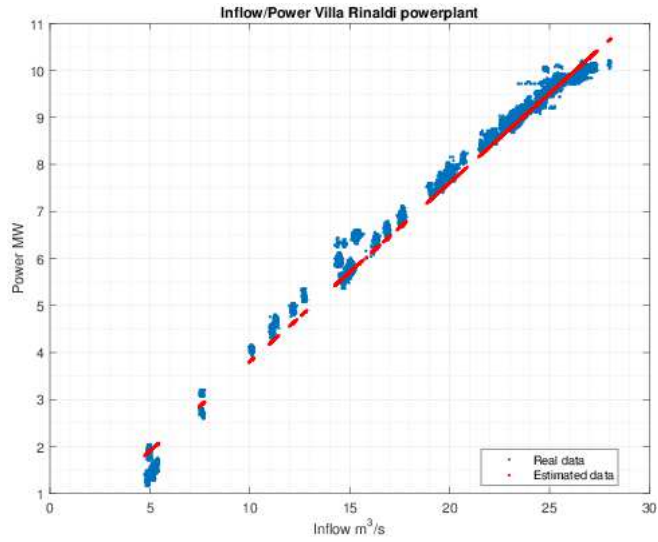}
	\label{Linear_caracterization}
	\caption{Real power v.s. estimated power by a linear model.}
\end{figure}

	A reasonable question could be: is the linear caracterization correct? What is the error committed approximating the most general low:
\begin{equation}
E=\mathcal{E} \left(\boldsymbol{\theta} \right), \quad \mathcal{E}: \mathbb{R}^n \to \mathbb{R},
\nonumber	
\end{equation}
whit a linear-one like:
\begin{equation}
\mathcal{E}=K x\left(t\right)
\nonumber
\end{equation}
where $x\left(t \right)$ is the flow and $K$ is the conversion coefficient from water to power?
Let's have a look to Figure $\eqref{Linear_caracterization}$ where on the $x$-axe we have the inflow and on the $y$-axe we have the corresponding power: this a typical Francis turbine graphic.

We observe that the error committed is little: the maximum error is $0.97 MW$ and it is committed when the inflow $x\left( t \right)$ is low. As observed in the article, due to the conformation of the problem, a power plant stay off or produces at maximum power: for this reason we can say that the error committed is acceptable if we deal with a Francis turbine.
\par
\begin{figure}[!ht]
	\centering
	\includegraphics[scale=0.5]{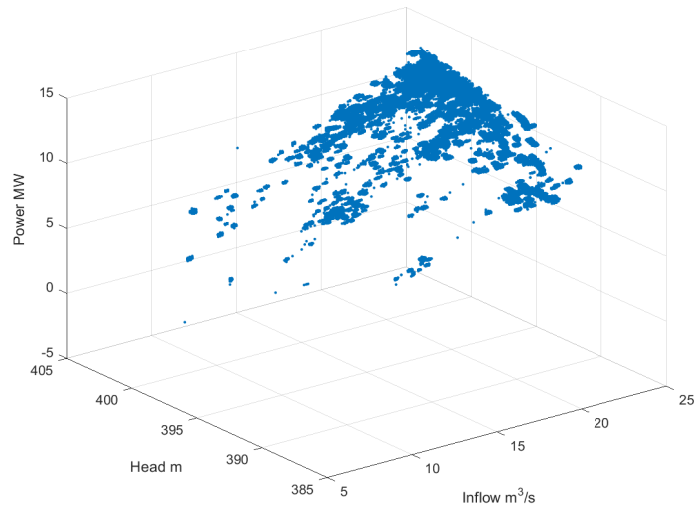}
	\label{NonLinear_caracterization}
	\caption{Real data: you can see how the generated power depends on the amount of water turbined and on the head.}
\end{figure}

 What happens if you deal with a Kaplan turbine? Remember that its invention allowed efficient power production in low-head applications that was not possible with Francis turbines.
Is it still correct to approximate the function $E=\mathcal{E} \left(\boldsymbol{\theta} \right)$ whit a linear one? Let's have a look to the Figure $\eqref{NonLinear_caracterization}$.

\begin{figure}[!ht]
	\centering
	\includegraphics[scale=0.5]{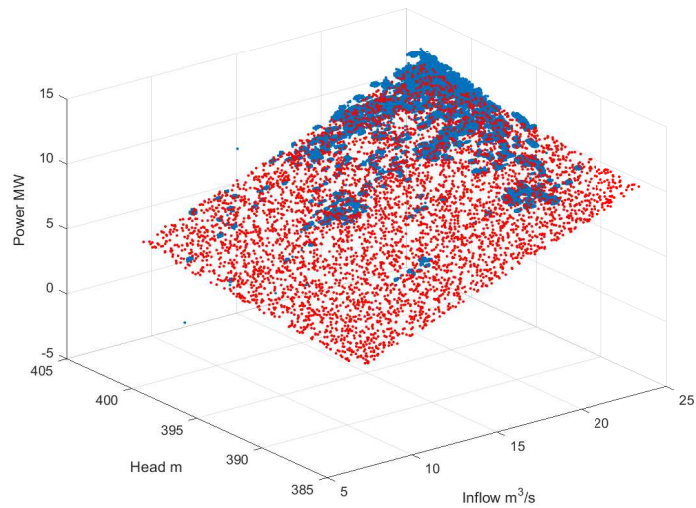}
	\label{Linear_Fitted_data_Head_dependence}
	\caption{Fitted data: a check of how the model proposed could consider the head dependance.}
\end{figure}

It is clear that in this case you can't use a simple linear model. As fist approximation you can use a model like:
\begin{equation}
	\mathcal{E}=\beta_{0} + \beta_{1} x\left(t\right) + \gamma_{0} + \gamma_{1} h\left(t\right)x\left(t\right),
\nonumber
\end{equation}
where $x$ is the flow and $h$ is the head. It is obvious that, for a physical reason, $\beta_{0}=0$ and $\gamma_{0}=0$. The problem using such a expression to calculate the power is that a non-linearity arises: indeed $h\left(t\right)=h\left(x\left(t\right)\right)$, since the head $h=h\left(V\left(t\right) \right)$ and $V\left(t\right)=x\left(t\right)$, and so you have a quadratic dependence on $x$. You can solve this problem using a form for $\mathcal{E}$ like:
\begin{equation}
\mathcal{E}=\beta_{0} + \beta_{1} x\left(t\right) + \gamma_{1} h\left(t\right),
\nonumber
\end{equation}
In this case the problem remains linear but you have to calculate the head $h\left(t\right)$, i.e. the volume $V\left(t\right)$, for each time $t$. In Figure $\eqref{Linear_Fitted_data_Head_dependence}$ it is shown how this formulation for $\mathcal{E}$ can take into account the power head dependence \footnote{All this models were created using $fitlm$ function avaiable in $Matlab$. }.

\clearpage
	
\end{appendices}

\bibliographystyle{plain}
\bibliography{Bibliografia}

\end{document}